# Interference phenomenon in electron-molecule collisions

A. S. Baltenkov[1] and I. Woiciechowski[2]

[1]*Arifov Institute of Ion-Plasma and Laser Technologies, 100125, Tashkent, Uzbekistan*
[2]*Alderson Broaddus University, 101 College Hill Drive, WV 26416, Philippi, USA*

**Abstract:** The article discusses how the pattern of elastic scattering of an electron on a pair of identical atomic spheres will look if we abandon the standard in the molecular physics assumption that, outside the molecular sphere, in the external region, the wave function of the molecular continuum is atomic-like and is a linear combination of regular and irregular solutions of the wave equations. To do this, the elastic scattering of slow particles by pair of non-overlapping short-range potentials has been studied. The continuum wave function of particle is represented as a combination of a plane wave and two spherical s-waves that propagated freely throughout space. The asymptotic of this function determines in closed form the amplitude of elastic particle scattering. It has been shown that this amplitude can be presented as a partial expansion in a set of ortho-normal functions $Z_\lambda(\mathbf{r})$ other, than spherical harmonics $Y_{lm}(\mathbf{r})$. General formulas for these functions are obtained. The coefficients of the scattering amplitude expansion to a series of functions $Z_\lambda(\mathbf{r})$ determine the scattering phases $\eta_\lambda(k)$ for the two-atomic target under consideration. The special features of the S-matrix method for the case of arbitrary non-spherical potentials are discussed.

**Keywords:** Electron elastic scattering, molecular sphere, matching conditions, S-matrix method

**Introduction**

In the molecular collision physics, the idea that *molecular continuum electron functions are similar to those for electron-atom scattering* (*) [1], are considered as a matter-of-course and, as far as we know, they are beyond any doubt. It cannot be said that the statement (*) is obvious. In the muffin-tin-model of molecule, the latter is represented as a cluster of N non-overlapping atomic spheres. A classical physical picture of wave scattering on such a target is based on the Huygens-Fresnel principle. According to this principle, the initial wave interacts with each target center that becomes a source of the secondary spherical scattered waves and beyond the target there is N the spherical waves outgoing from each of the centers, rather than a set of partial spherical waves having the asymptotic form

$$[R_{kl}(r)Y_{lm}(\mathbf{r})]_{r\to\infty} \approx e^{i(\eta_l+\frac{\pi l}{2})}\frac{1}{kr}\sin(kr-\frac{\pi l}{2}+\eta_l)Y_{lm}(\mathbf{r}), \qquad (1)$$

as is the case in the scattering of an electron by an atom. In equation (1) **r** is the radius-vector of scattered electron relative to the center of mass of the molecule; **k** is the electron wave vector. Here and throughout what follows, we use the atomic system of units.

Apparently, the statement (*) is related to standard methods for calculating the molecular continuum. According to reviews [1-3], the configuration space describing the scattered or ejected electron and the molecule is divided into two regions: internal and external. Some sphere with radius $r_m$ defines an internal region in which the main part of the charge distribution of the molecular states of interest is located. The electron continuum wave functions in the external region represented as a linear combination of the regular and irregular solutions of the Schrödinger equation for the potential that in this region "*is taken to be spherical ... about the molecular center*" [4]. A microscopic observer (Maxwell's demon), located inside the sphere, sees a picture of the scattering of an electron wave at several centers. But when he moves outside the sphere, he observes the pattern of wave scattering on a single center located in the center of mass of the molecule. What will be the scattering pattern if we abandon the assumption written above by the italics?

A qualitative picture of the scattering of an electron wave by a diatomic target is shown in Figure 1(a, b). On the left in the figure is an electron plane wave, and on the right are two spherical waves, as follows from the Huygens-Fresnel principle. The interference of these waves creates a diffraction scattering pattern whose properties depend periodically on the ratio of the inter-atomic distance to the electron wavelength. Figure 1(b) allows you to see how the electron wave scattering pattern is transformed after the introduction of the sphere $r_m$. Outside molecular sphere, there is a single spherical wave centered at the center of the sphere but not two waves, as in Figure 1(a).

In the connection with this Figure a question arises: Is it possible to adapt the method of partial waves for the case of a multicenter target keeping the Huygens-Fresnel picture of the scattering process according to which far from the target there is a system of the secondary waves outgoing from each of the centers? Demkov and Rudakov gave the positive answer to this question in paper [5] where it was shown that the S-matrix method could be also applied to non-spherical potentials. In the present paper as a simple example of scattering of a slow particle by two short-range non-overlapping potentials we will analyze the special features of the partial wave method for non-spherical targets. We will consider the targets with fixed positions of atoms in space.

The outline of our article is as follows. In Section 1, we will calculate the scattering amplitude of a particle on a pair of short-range potentials based on the scattering pattern presented in Figure 1(a). It will be shown that this amplitude can be written in a closed form, but not as an expansion in partial waves. In Section 2, we describe the partial wave method for the scattering pattern Figure 1(a). Expansion of the scattering amplitude in a series of functions $Z_\lambda(\mathbf{k})$ is carried out in Section 3. Section 4 contains some numerical calculations. Section 5 is Conclusions.

**1. Elastic scattering of a particle by two short-range potentials**

Based on Brueckner's article [6], we will follow how the problem of particle scattering by a pair of non-overlapping short-range potential wells with radius $\rho$ is solved. In the coordinate system with the origin at the center of mass of pair of identical potential wells the continuum wave function, according to Figure 1(a), is represented as a combination of a plane wave and two spherical s-waves, generated by short-range potentials with centers at the points $\mathbf{r} = \pm \mathbf{R}/2$

$$\psi_{\mathbf{k}}^+(\mathbf{r}) = e^{i\mathbf{k}\cdot\mathbf{r}} + D_1(\mathbf{k})\frac{e^{ik|\mathbf{r}+\mathbf{R}/2|}}{|\mathbf{r}+\mathbf{R}/2|} + D_2(\mathbf{k})\frac{e^{ik|\mathbf{r}-\mathbf{R}/2|}}{|\mathbf{r}-\mathbf{R}/2|} ; \text{ for } |\mathbf{r}\pm\mathbf{R}/2| > \rho. \qquad (2)$$

Function (2) is the general solution to the wave equation outside the scatterer's spheres with radius $\rho$ [6].

The so far unknown coefficients $D_1(\mathbf{k})$ and $D_2(\mathbf{k})$ in (2) are found as a result of imposing boundary conditions on trial wave function $\psi_{\mathbf{k}}^+(\mathbf{r})$ at the points $\mathbf{R}/2$ and $-\mathbf{R}/2$. Let's find these boundary conditions. To do this, we study the behavior of a particle in the s-state in the field of a short-range spherical potential well. Outside the well at a distance larger than the range of potential $\rho$, the following particle wave function satisfies the wave equation

$$\psi_0(\mathbf{r}) \propto \frac{1}{kr}\sin(kr+\delta_0)Y_{00}(\mathbf{r}). \qquad (3)$$

Here and further $Y_{lm}(\mathbf{r}) \equiv Y_{lm}(\vartheta_r,\varphi_r)$; $\vartheta_r$ and $\varphi_r$ are spherical angles of the vector $\mathbf{r}$. Let us assume that we know the s-phase of the elastic scattering of a particle on an isolated potential well, that is, the function $\delta_o(k)$ over the entire range of the wave vector $k$. Passing to the limit $r \to 0$ in formula (3), we obtain the following boundary conditions imposed on function (2) at the centers of potential well

$$\psi_0(\mathbf{r})_{r\to 0} \propto C\left[\frac{1}{r} + k\cot\delta_0(k)\right]. \qquad (4)$$

For wells centered at points $\mathbf{r} = \pm\mathbf{R}/2$, we obtain the following boundary conditions for function (2)

$$\psi_{\mathbf{k}}^+(\mathbf{r})_{\mathbf{r}\to\mathbf{R}/2} \approx C_1\left[\frac{1}{|\mathbf{r}-\mathbf{R}/2|} + k\cot\delta_0(k)\right];$$
$$\psi_{\mathbf{k}}^+(\mathbf{r})_{\mathbf{r}\to-\mathbf{R}/2} \approx C_2\left[\frac{1}{|\mathbf{r}+\mathbf{R}/2|} + k\cot\delta_0(k)\right]. \qquad (5)$$

$C_1$ and $C_2$ here are the some constants. Applying formulas (5) to the function (2) we obtain the *exact general solution* of the Schrödinger equation outside the scatterer's spheres that describes the multiple scattering of particle by two-center target [6]. The amplitude of particle scattering by the target is obtained by considering the asymptotic behavior of the wave function (2). As result, we obtain the following *exact amplitude* of the particle multiple s-scattering by given two-center target [6, 7]

$$F(\mathbf{k},\mathbf{k}',\mathbf{R}) = \frac{2}{a^2-b^2}\{b\cos[(\mathbf{k}-\mathbf{k}')\cdot\frac{\mathbf{R}}{2}] - a\cos[(\mathbf{k}+\mathbf{k}')\cdot\frac{\mathbf{R}}{2}]\}. \qquad (6)$$

We not reproduced here the details, but refer the reader to the original paper [7]. As in [7], we use here the following notation: $\mathbf{k}$ and $\mathbf{k}'$ are the particle linear momentums before and after scattering, respectively; the functions $a = \exp(ikR)/R$ and $b = k(i - \cot\delta_0) = -1/f_0(k)$. Here $f_0(k)$ is the *s*-partial amplitude of particle elastic scattering by isolated potential well. The total cross section of particle scattering in [6] is obtained from the amplitude (6) using the optical theorem [8]

$$\sigma(\mathbf{k},\mathbf{R}) = \int \frac{d\sigma}{d\Omega_{k'}} d\Omega_{k'} = \frac{4\pi}{k} \operatorname{Im} F(\mathbf{k}=\mathbf{k}',\mathbf{R}) = \frac{8\pi}{k} \operatorname{Im}\left[\frac{b - a\cos(\mathbf{k}\cdot\mathbf{R})}{a^2 - b^2}\right]. \qquad (7)$$

The case when a pair of non-identical atomic spheres forms the target was considered in [9]. The generalization of the formulas of this section to a larger number of scattering centers is not difficult. The coefficients for spherical s-waves are obtained in this case from the boundary conditions (5) imposed at the centers of all atomic spheres forming the target. The application of the Huang-Yang multipolar pseudopotential [10-12] for each of the scattering centers makes it possible to generalize the results presented here for the s-wave to the case of spherical waves with nonzero orbital angular momentum.

### 3. Method of partial waves for non-spherical targets

A molecular potential as a cluster of non-overlapping atomic potentials centered at the atomic sites is non-spherical. The solution $\psi_{\mathbf{k}}^{+}(\mathbf{r})$ of the Schrödinger equation with this potential is impossible to present at an arbitrary point of space as an expansion in spherical functions $Y_{lm}(\mathbf{r})$. However, asymptotically at great distances from the molecule, according to Demkov and Rudakov article [5], the continuum wave function of electron can be presented as an expansion in a set of other than $\psi_{\mathbf{k}}^{+}(\mathbf{r})$ orthonormal functions $Z_\lambda(\mathbf{r})$ and $Z_\lambda(\mathbf{k})$:

$$\psi_{\mathbf{k}}^{+}(\mathbf{r}\to\infty) \approx 4\pi \sum_\lambda R_{k\lambda}(r) Z_\lambda(\mathbf{r}) Z_\lambda^*(\mathbf{k}) \qquad (8)$$

with the partial wave function, as opposed to (1), determined by the following formula

$$[R_{k\lambda}(r) Z_\lambda(\mathbf{r})]_{r\to\infty} \approx e^{i(\eta_\lambda + \frac{\pi}{2}\omega_\lambda)} \frac{1}{kr} \sin(kr - \frac{\pi}{2}\omega_\lambda + \eta_\lambda) Z_\lambda(\mathbf{r}). \qquad (9)$$

Here the index $\lambda$ enumerates different partial wave functions similar to the quantum numbers $l$ and $m$ for the central field; $\omega_\lambda$ is the quantum number that is equal to the orbital momentum $l$ for the spherical symmetry case; $\eta_\lambda(k)$ are the "proper molecular phases". The explicit form of functions $Z_\lambda(\mathbf{k})$ (in terminology [5] "characteristic amplitudes") depends upon the specific type of the target field, particularly on the number of atoms forming the target and on mutual disposition of the scattering centers in space, *etc*. Functions $Z_\lambda(\mathbf{k})$, like the spherical functions $Y_{lm}(\mathbf{k})$, create an orthonormal system and for this reason:

$$\int Z_\lambda^*(\mathbf{k}) Z_\mu(\mathbf{k}) d\Omega_k = \delta_{\lambda\mu}. \qquad (10)$$

The elastic-scattering amplitude for a non-spherical target, according to [5], is given by the following expression

$$F(\mathbf{k},\mathbf{k'}) = \frac{2\pi}{ik}\sum_\lambda (e^{2i\eta_\lambda} - 1)Z_\lambda^*(\mathbf{k})Z_\lambda(\mathbf{k'}). \tag{11}$$

The total elastic scattering cross section, i.e. the cross section integrated over all directions of momentum of the scattered electron **k'**, is defined by the formula

$$\sigma(\mathbf{k}) = \frac{(4\pi)^2}{k^2}\sum_\lambda |Z_\lambda(\mathbf{k})|^2 \sin^2 \eta_\lambda(k). \tag{12}$$

Of course, this cross section depends on the mutual orientation of incident electron momentum **k** and molecule axes. The cross section averaged over all the directions of momentum of incident electron **k** is connected with the molecular phases $\eta_\lambda(k)$ by the following formula

$$\bar{\sigma}(k) = \frac{4\pi}{k^2}\sum_\lambda \sin^2 \eta_\lambda(k). \tag{13}$$

In the case of a spherical symmetrical target the formula (13) exactly coincides with the known formula for the total scattering cross section. Indeed, in the case of the central field the index $\lambda$ is replaced by the quantum numbers $l$ and $m$. But the phase of scattering by the central field is independent of the magnetic number and therefore for the given value of the orbital momentum $l$ it is necessary to summarize over all $m$. This results in the factor $(2l+1)$ under the sum sign in formula (13).

The partial wave (9) and molecular phases $\eta_\lambda(k)$ are classified, according to [5], by their behavior for low electron energies, i.e. for $k \to 0$. In this limit the particle wavelength is great as compared with the target size and the function $Z_\lambda(\mathbf{k})$ tends to some spherical function $Y_{lm}(\mathbf{k})$. The corresponding phase is characterized in this limit by the following asymptotic behavior: $\eta_\lambda(k) \to k^{2\lambda+1}$.

For the molecular system that is created by two short-range potentials each of which is a source of the scattered s-waves, the molecular phase shifts $\eta_\lambda(k)$ and the functions $Z_\lambda(\mathbf{k})$ can be calculated in the explicit form (see [7] and [13] in case of low electron energies). This simplest multicenter system is a good example illustrating the method of partial waves for non-spherical targets formed by non-overlapping atomic potentials. It is important that the problem of slow particle scattering by this system of centers can be solved analytically.

**4. Expansion of the scattering amplitude (4) in a series of functions $Z_\lambda(\mathbf{k})$**

According to [5], the exact amplitude (6) should be represented as a partial expansion in a series of functions $Z_\lambda(\mathbf{k})$. Following the papers [7, 14] let us rewrite the scattering amplitude (6) in following form

$$F(\mathbf{k},\mathbf{k'},\mathbf{R}) = -\frac{2}{a+b}\cos(\mathbf{k}\cdot\mathbf{R}/2)\cos(\mathbf{k'}\cdot\mathbf{R}/2) + \frac{2}{a-b}\sin(\mathbf{k}\cdot\mathbf{R}/2)\sin(\mathbf{k'}\cdot\mathbf{R}/2). \tag{14}$$

The amplitude (14) should be considered as a sum of two partial amplitudes. The first of them is written as

$$\frac{4\pi}{2ik}(e^{2i\eta_0}-1)Z_0(\mathbf{k})Z_0^*(\mathbf{k}') = -\frac{2}{a+b}\cos(\mathbf{k}\cdot\mathbf{R}/2)\cos(\mathbf{k}'\cdot\mathbf{R}/2).\tag{15}$$

The second one is defined by the following expression

$$\frac{4\pi}{2ik}(e^{2i\eta_1}-1)Z_1(\mathbf{k})Z_1^*(\mathbf{k}') = \frac{2}{a-b}\sin(\mathbf{k}\cdot\mathbf{R}/2)\sin(\mathbf{k}'\cdot\mathbf{R}/2).\tag{16}$$

After elementary transformations of (15) and (16), we obtain two molecular phases

$$\cot\eta_0 = -\frac{qR+\cos kR}{kR+\sin kR},\qquad \cot\eta_1 = -\frac{qR-\cos kR}{kR-\sin kR}.\tag{17}$$

Comparing these expressions with the general formula (9) we come to the following conclusions. If the electron states are characterized by a projection of the angular momentum on the $\mathbf{R}$ axis and by parity of the wave function relative to the reflection in the plane perpendicular to $\mathbf{R}$ and going through the middle of the inter-atomic distance, then the first of the partial waves (17) corresponds to the state $\Sigma_g$ and the second one to $\Pi_u$.

Here the wave number $q(k) = -k\cot\delta_0(k)$. The molecular phases $\eta_\lambda(k)$ in (17) can be classified by considering their behavior at $k\to 0$ [5]. In this limit the electron wavelength is much greater than the target size and the picture of scattering should approach the case of spherical symmetry. Considering the transition to this limit in the formulas (17), we obtain: $\eta_0(k\to 0)\sim k$ and $\eta_1(k\to 0)\sim k^3$. Thus, the molecular phases behave similar to the $s$- and $p$- phases in the spherically symmetrical potential, which explains the choice of their indexes.

In paper [13], formulas for molecular phases (17) were obtained in the limit of low electron energies, for which the electron scattering length on the A atom forming the target is related to the wave vector $q(k)$ as follows $1/A = [-k\cot\delta_0(k)]_{k\to 0}$. Let us establish a relation between the molecular electron scattering length $A_m$ and the electron scattering length on the atoms $A$. These scattering lengths are determined by the following formulas [14]

$$1/A_m = [-k\cot\eta_0(k)]_{k\to 0} \text{ and } 1/A = [-k\cot\delta_0(k)]_{k\to 0}.\tag{18}$$

Multiplying both parts of the zeroes phase (17) by and passing to the limit $k\to 0$, we obtain the required relation between the scattering lengths (compare with [15])

$$\frac{1}{A_m} = \frac{R+A}{2RA}.\tag{19}$$

From formulas (15) and (16) we obtain two "characteristic amplitudes"

$$Z_0(\mathbf{k}) = \frac{\cos(\mathbf{k} \cdot \mathbf{R}/2)}{\sqrt{2\pi S_+}}, \qquad Z_1(\mathbf{k}) = \frac{\sin(\mathbf{k} \cdot \mathbf{R}/2)}{\sqrt{2\pi S_-}}. \qquad (20)$$

Here $S_\pm = 1 \pm j_0(kR)$. Function $j_0(kR)$ here is the spherical Bessel function [16]. It is easy to demonstrate that the functions $Z_\lambda(\mathbf{k})$, like the spherical functions $Y_{lm}(\mathbf{k})$, create an orthonormal system. The functions (20) are defined by the geometrical target structure, i.e. by the direction of the molecular axis $\mathbf{R}$ in the arbitrary coordinate system, in which the electron momentum vectors before and after scattering are $\mathbf{k}$ and $\mathbf{k'}$, respectively. The transition to the limit $k \to 0$ in formulas (17) gives instead of the functions $Z_\lambda(\mathbf{k})$ the well-known spherical functions

$$Z_0(\mathbf{k})_{k \to 0} \to \frac{1}{\sqrt{4\pi}} \equiv Y_{00}(\mathbf{k}), \qquad Z_1(\mathbf{k})_{k \to 0} \to \sqrt{\frac{3}{4\pi}} \cos\vartheta \equiv Y_{10}(\mathbf{k}). \qquad (21)$$

Here $\vartheta$ is the angle between the vector $\mathbf{k}$ and axis $\mathbf{R}$. So, *only in the limit $k \to 0$ it becomes correct to replace a non-spherical molecular field in external region (beyond the molecular sphere) by a spherical one*.

Summarize the results obtained with the method of partial waves for a target formed by non-overlapping atomic potentials. For two atomic targets the molecular phases of scattering $\eta_\lambda(k)$ and the functions $Z_\lambda(\mathbf{k})$ can be found explicitly. The phases of molecular scattering, as the atomic ones, are the functions of electron momentum $k = |\mathbf{k}|$ only. The form of the functions $Z_\lambda(\mathbf{k})$ is defined by the structure of a target and its orientation in space. The number of non-zero molecular phases in this case is equal to two. This is connected with the fact that each of these two scattering centers is a source of s-spherical waves only, which is valid for the case of low electron energy. If the scattering by each of these centers would be accompanied by generation of spherical waves with non-zero orbital momenta (this case was considered in paper [17]) then the number of non-zero molecular phases $\eta_\lambda(k)$ would be greater.

Other generalizations of spherical harmonics are well known. For example, prolate spheroidal and Coulomb spheroidal angular functions [18] that are used for description of an electron behavior in the field of a diatomic-molecule cations. Or the so-called dipole functions [19] that are convenient to use if the target-molecule has a constant dipole moment.

## 5. Numerical results

To illustrate the potentialities of the formulas obtained in Section 4, we calculate functions (20) and differential cross sections for electron scattering (12) on a model quasi-molecule formed by a pair of identical atomic hard-spheres for which s-wave phase shifts is [19]

$$\delta_0(k) = -k\rho, \qquad (22)$$

where $\rho$ is the radius of scatterer spheres in (2). The hard-sphere radius $\rho$ can also be looked upon as the scattering length $A$ (18). The distance between the centers of atomic spheres is considered equal to $R=1.094$ Å $=2.067$ atomic units (au); radii of atomic spheres $\rho = 1$ au. The behavior of the scattering phases as the functions of the electron momentum $k$ have been discussed by us (and before us) in a number of papers [5, 7, 9, 13, 14], while the special features of the $Z_\lambda$ functions have not been studied by anyone. It turned out that the behavior of the $Z_\lambda$ functions is rather nontrivial. Next, we will focus on the study of these functions. The some numerical results necessary for their calculation are given in the Table.

Table. All parameters in the Table are given in atomic units.

| $k$ | $\sin^2\eta_0(k)$, Eq. (17) | $\sin^2\eta_1(k)$, Eq. (17) | $\overline{\sigma}(k)$, Eq. (13) |
|---|---|---|---|
| 0.1 | 0.0180 | $0.1845 \cdot 10^{-05}$ | 22.6792 |
| 1.0 | 0.9230 | 0.3025 | 15.4009 |
| 2.0 | 0.6464 | 0.9317 | 4.9578 |
| 3.0 | 0.0203 | 0.0195 | 0.0556 |
| 4.0 | 0.6500 | 0.4872 | 0.8932 |
| 5.0 | 0.8712 | 0.9539 | 0.9174 |

The results of numerical calculations of "characteristic amplitudes" (20) [20] are presented in Figure 2. The angle $\theta$ in this figure is the angle between the vectors **k** and **R** for functions $Z_\lambda(\mathbf{k})$; for functions $Z_\lambda(\mathbf{r})$ $\theta$ is the angle between the vectors **r** and **R**. Dotted lines are the spherical harmonics $Y_{00}(\cos\vartheta)$ and $Y_{10}(\cos\vartheta)$ to which the calculated curves with indices $k=1.0, 2.0 \ldots$ tend in the limit $k \to 0$.

Figure 3 shows the partial differential cross sections (12)

$$\sigma_0(\mathbf{k}) = \frac{(4\pi)^2}{k^2} |Z_0(\mathbf{k})|^2 \sin^2\eta_0(k);$$

$$\sigma_1(\mathbf{k}) = \frac{(4\pi)^2}{k^2} |Z_1(\mathbf{k})|^2 \sin^2\eta_1(k), \tag{23}$$

and the sum of these partial sections $\sigma(\mathbf{k})$ (12) as the functions of the angle between the vectors **R** and **k**. The **R** axis of the molecule is the polar axis in these figures. 3D-pictures of differential cross sections are figures of rotation of curves around the **R**-axis. Dotted lines - the summa of partial cross sections - do not coincide with circles, as it would be in the case of spherical waves (1). The passage of the curves through the center of the polar coordinate system is associated with the zeroes of the curves in Figure 2, and the more these zeros, the more lobes on the curves of the cross sections $\sigma_0(\mathbf{k})$ and $\sigma_1(\mathbf{k})$. Numerical calculations of the cross section $\sigma(\mathbf{k}, \mathbf{R})$ using the optical theorem, that is, according to formula (7), as expected, coincide with the total cross sections presented in Figure 3 by dotted lines.

It is simple to calculate the functions $Z_\lambda(\mathbf{k})$ and the scattering phases $\eta_\lambda(k)$ when one knows the exact wave function (2). On the other hand, if this function is known, as it is the case for a system of non-overlapping potentials, the scattering amplitude can be obtained in the closed form (6) and the cross section can be found with the help of the optical theorem (7), and therefore there is no necessity to resort to the method of partial waves. However, for non-spherical potentials different from muffin-tin-potentials, i.e. when the model of non-overlapping potential becomes inapplicable, the use of the partial wave method [5] makes it possible to separate in the explicit form the scattering dynamics contained in the molecular phases $\eta_\lambda(k)$ from the kinematics of the process defined by the functions $Z_\lambda(\mathbf{k})$.

**5. Conclusions**

In studies of electron-molecule scattering, the continuum electron is usually treated [1-4] as moving in a spherically averaged molecular field [see figure 1(b)]. The wave functions describing the scattering of an electron by a polyatomic molecule outside the so-called molecular sphere (in the external region) are considered as a linear combination of regular and irregular solutions of the wave equation. The phase shifts of molecular wave function are defined from the matching condition for the solutions of the wave equation inside and beyond this sphere. Obviously, the introduction of a molecular sphere changes the scattering pattern [compare Fig. 1(a) and (b)]. The diffraction pattern of the electron wave itself disappears, since there is only one spherical wave far from the target. Whereas terms $\sin kR/kR$ in formulas (17), obtained within the scattering pattern 1(a), there is a clear manifestation of the electron diffraction of a pair of spherical s-waves emitted by spatially separated sources. Apparently, the traditional approach based on formulas (1) and (2) and Fig. (b) as a result allows one to obtain the scattering phases of a particle on an isolated molecular sphere, rather than on a real two-center target.

Thus, the statement (*) that the *molecular continuum electron functions are similar to those for electron-atom scattering* should be supplemented with the words: *if we assume that in the external region the continuum wave function is atomic-like.*

In the present article, we abandoned this assumption and allowed spherical waves to propagate freely throughout space and come to a different result. Instead of partial spherical waves (1), obtained within the standard methods for calculating the molecular continuum, we have the following pair of wave functions (9)

$$[R_{k0}(r)Z_0(\mathbf{r})]_{r\to\infty} \approx e^{i\eta_0}\frac{1}{kr}\sin(kr+\eta_0)Z_0(\mathbf{r});$$

$$[R_{k1}(r)Z_1(\mathbf{r})]_{r\to\infty} \approx e^{i(\eta_1+\frac{\pi}{2})}\frac{1}{kr}\sin(kr-\frac{\pi}{2}+\eta_1)Z_1(\mathbf{r}). \qquad (24)$$

Since in these asymptotics of the wave function of the continuum instead of spherical harmonics $Y_{lm}(\mathbf{r})$ there are functions $Z_\lambda(\mathbf{r})$, it becomes clear that: *no matter how far we move away from the non-spherical target, we will never be able to consider the target as atomic-like, and the wave function of the particle consider as the spherical waves*; except the case when the particle wavelength is much more than the target size $1/k>>R$.

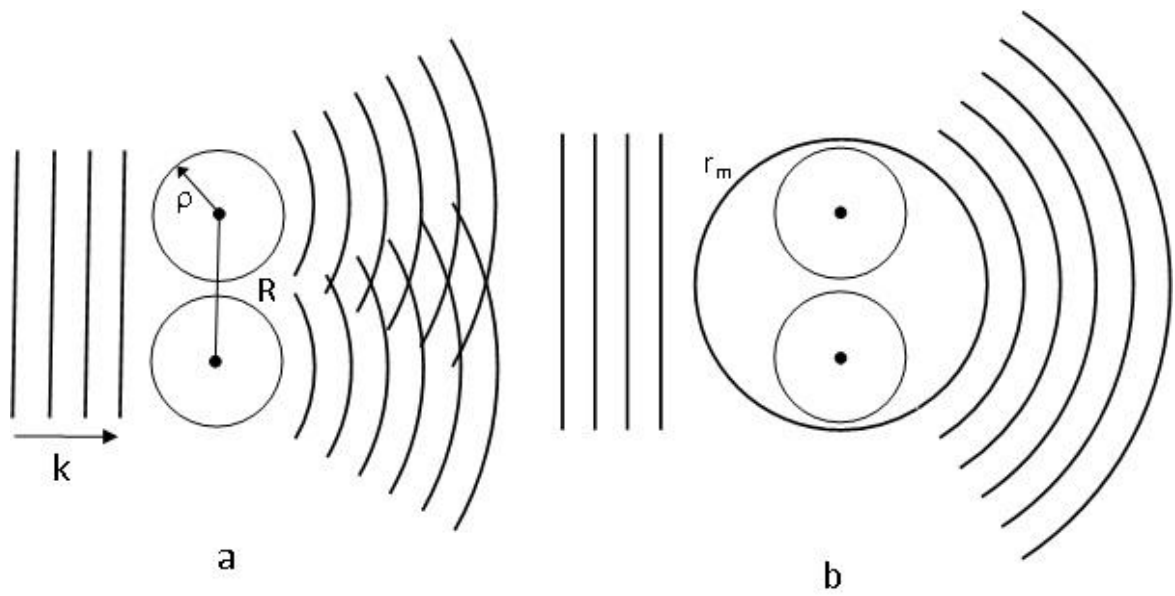

Figure 1. A qualitative picture of the scattering of an electron wave by a diatomic target; (a) is the Huygens-Fresnel picture; (b) is the Dill and Dehmer picture [4].

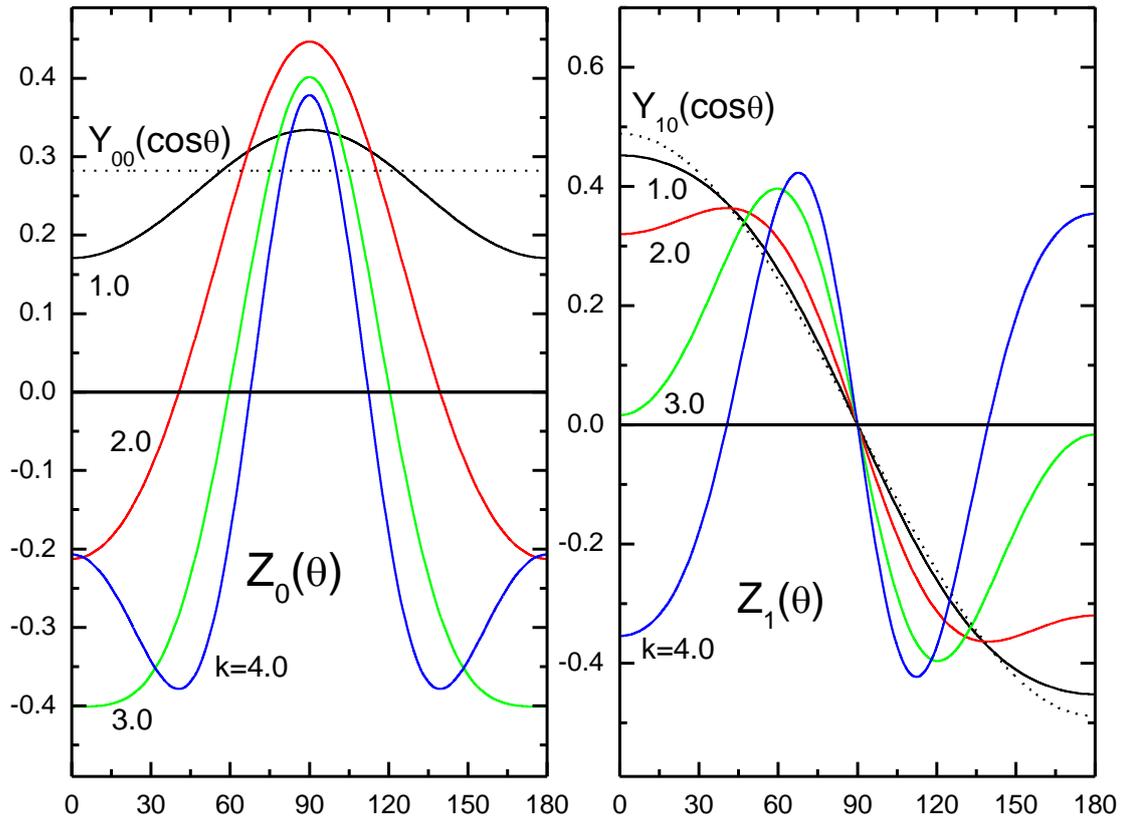

Figure 2. The "characteristic amplitudes" (20) as the functions of angle $\theta$.

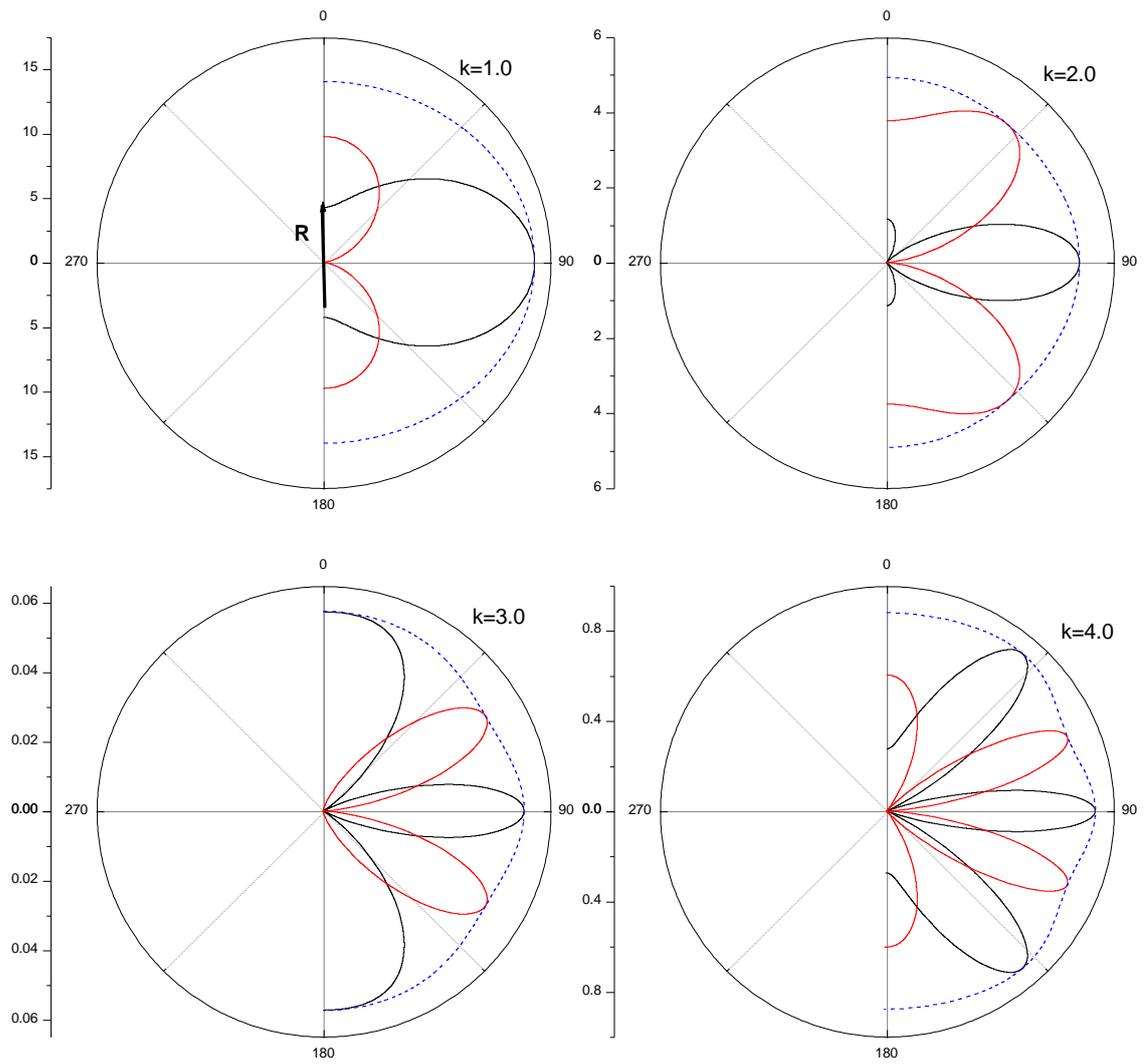

Figure 3. The partial cross sections $\sigma_0(\mathbf{k})$ and $\sigma_1(\mathbf{k})$ (23), and their summa (12) as the functions of the angle between the vectors $\mathbf{R}$ and $\mathbf{k}$ for fixed values of *k*. Solid black lines are $\sigma_0(\mathbf{k})$; solid red lines are $\sigma_1(\mathbf{k})$; dotted lines are cross sections (12).

13